\documentclass[amsmath,longbibliography,superscriptaddress,reprint]{revtex4-2}
\pdfoutput=1
\usepackage{hyperref}
\usepackage{graphicx}
\usepackage{siunitx}

\begin{document}
	
	\title{Formation of Organic Color Centers in Air-Suspended Carbon Nanotubes Using Vapor-Phase Reaction}

	\author{Daichi~Kozawa}
	\email[Corresponding author. ]{daichi.kozawa@riken.jp}	
	\affiliation{Quantum Optoelectronics Research Team, RIKEN Center for Advanced Photonics, Saitama 351-0198, Japan}

	\author{Xiaojian~Wu}
	\affiliation{Department of Chemistry and Biochemistry, University of Maryland, College Park, Maryland 20742, United States}

	\author{Akihiro~Ishii}
	\affiliation{Quantum Optoelectronics Research Team, RIKEN Center for Advanced Photonics, Saitama 351-0198, Japan}
	\affiliation{Nanoscale Quantum Photonics Laboratory, RIKEN Cluster for Pioneering Research, Saitama 351-0198, Japan}

	\author{Jacob~Fortner}
	\affiliation{Department of Chemistry and Biochemistry, University of Maryland, College Park, Maryland 20742, United States}

	\author{Keigo~Otsuka}
	\affiliation{Nanoscale Quantum Photonics Laboratory, RIKEN Cluster for Pioneering Research, Saitama 351-0198, Japan}

	\author{Rong~Xiang}
	\affiliation{Department of Mechanical Engineering, The University of Tokyo, Tokyo 113-8656, Japan}

	\author{Taiki~Inoue}
	\affiliation{Department of Mechanical Engineering, The University of Tokyo, Tokyo 113-8656, Japan}

	\author{Shigeo~Maruyama}
	\affiliation{Department of Mechanical Engineering, The University of Tokyo, Tokyo 113-8656, Japan}

	\author{YuHuang~Wang}
	\affiliation{Department of Chemistry and Biochemistry, University of Maryland, College Park, Maryland 20742, United States}
	\affiliation{Maryland NanoCenter, University of Maryland, College Park, Maryland 20742, United States}

	\author{Yuichiro~K.~Kato}
	\email[Corresponding author. ]{yuichiro.kato@riken.jp}
	\affiliation{Quantum Optoelectronics Research Team, RIKEN Center for Advanced Photonics, Saitama 351-0198, Japan}
	\affiliation{Nanoscale Quantum Photonics Laboratory, RIKEN Cluster for Pioneering Research, Saitama 351-0198, Japan}

	\begin{abstract}
		Organic color centers in single-walled carbon nanotubes have demonstrated exceptional ability to generate single photons at room temperature in the telecom range. Combining the color centers with pristine air-suspended tubes would be desirable for improved performance, but all current synthetic methods occur in solution which makes them incompatible. Here we demonstrate formation of color centers in air-suspended nanotubes using vapor-phase reaction. Functionalization is directly verified on the same nanotubes by photoluminescence spectroscopy, with unambiguous statistics from more than a few thousand individual nanotubes. The color centers show a strong diameter-dependent emission intensity, which can be explained with a theoretical model for chemical reactivity taking into account strain along the tube curvature. We are also able to estimate the defect density by comparing the experiments with simulations based on a one-dimensional diffusion equation, whereas the analysis of diameter dependent peak energies gives insight to the nature of the dopant states. Time-resolved measurements show a longer lifetime for color center emission compared to E$_{11}$ exciton states. Our results highlight the influence of the tube structure on vapor-phase reactivity and emission properties, providing guidelines for development of high-performance near-infrared quantum light sources.
	\end{abstract}

	\maketitle
		
	\section*{Introduction}	
	Quantum technologies offer various advantages beyond the classical limits in secure communications \cite{Gisin2007}, parallel computing \cite{DiVentra2013}, and sensing \cite{Degen2017}. Solid-state single-photon sources \cite{Aharonovich:2016} are a fundamental component in these technologies, and considerable progress has been made in various systems including quantum dots \cite{Michler:2000science}, diamond \cite{Kurtsiefer:2000}, SiC \cite{Castelletto:2014}, and two-dimensional materials \cite{Tran:2016}. Of practical interest are single-walled carbon nanotubes (SWCNTs), since operation at room temperature and in the telecom range is possible. In particular, organic color centers formed on nanotubes \cite{Brozena2019} offer additional advantages with their optical properties being chemically tunable using a variety of molecular precursors, including aryl-halides \cite{Kim2018, Gifford2019a, Wu2018, Kwon2016a}, diazonium-salt \cite{Iwamura2014, Hartmann:2016, Akizuki2015, Ishii:2018, Saha2018, Luo2019, Shiraki2017, Shiraki2016, Berger2019}, ozone \cite{Kim2016,Miyauchi:2013,Ghosh:2010Science}, and hypochlorite \cite{Lin2019} that can covalently bond to the carbon lattice. By introducing dopant states with different emission energies and achieving potential traps deeper than the thermal energy, single-photon sources with desired properties can be produced \cite{He:2017}. 
	
	Further development of quantum emitters with improved performance is expected if color centers can be introduced to as-grown air-suspended SWCNTs known for their pristine nature \cite{Moritsubo:2010, Ishii:2015}. These nanotubes exhibit bright photoluminescence (PL) because of a low quenching site density, which makes them ideal for single-photon sources. Existing methods, however, require liquid-phase reaction where solvents and surfactants will inevitably be in contact with the nanotubes, making them incompatible with air-suspended tubes. To combine the excellent optical properties of the air-suspended SWCNTs with these organic color centers, an intelligent design of chemical reaction is required.
	
	In this work, we propose and demonstrate a vapor-phase reaction to create organic color centers in air-suspended SWCNTs. Tubes are functionalized with a photochemical reaction where adsorbing precursor vapor allows for preserving the suspended structures because of a weak mechanical perturbation. Individual tubes are characterized by confocal microspectroscopy to verify the formation of color centers, and we conduct a statistical survey of more than 2000 PL spectra to investigate diameter-dependent emission intensities and energies. PL intensity changes are interpreted using a theoretical model for reactivity that considers strain along the curvature of a SWCNT. We are also able to estimate the defect density by comparing experimentally obtained quenching with simulations based on  numerical solutions of a diffusion equation. Characteristic trapping potential depths of color centers are studied by analyzing emission energies to elucidate the nature of the dopant states. Furthermore, we perform time-resolved PL measurements to investigate the color center emission and find that the functionalized tubes show a longer decay lifetime than the pristine tubes.
	
	\begin{figure*}
	\includegraphics{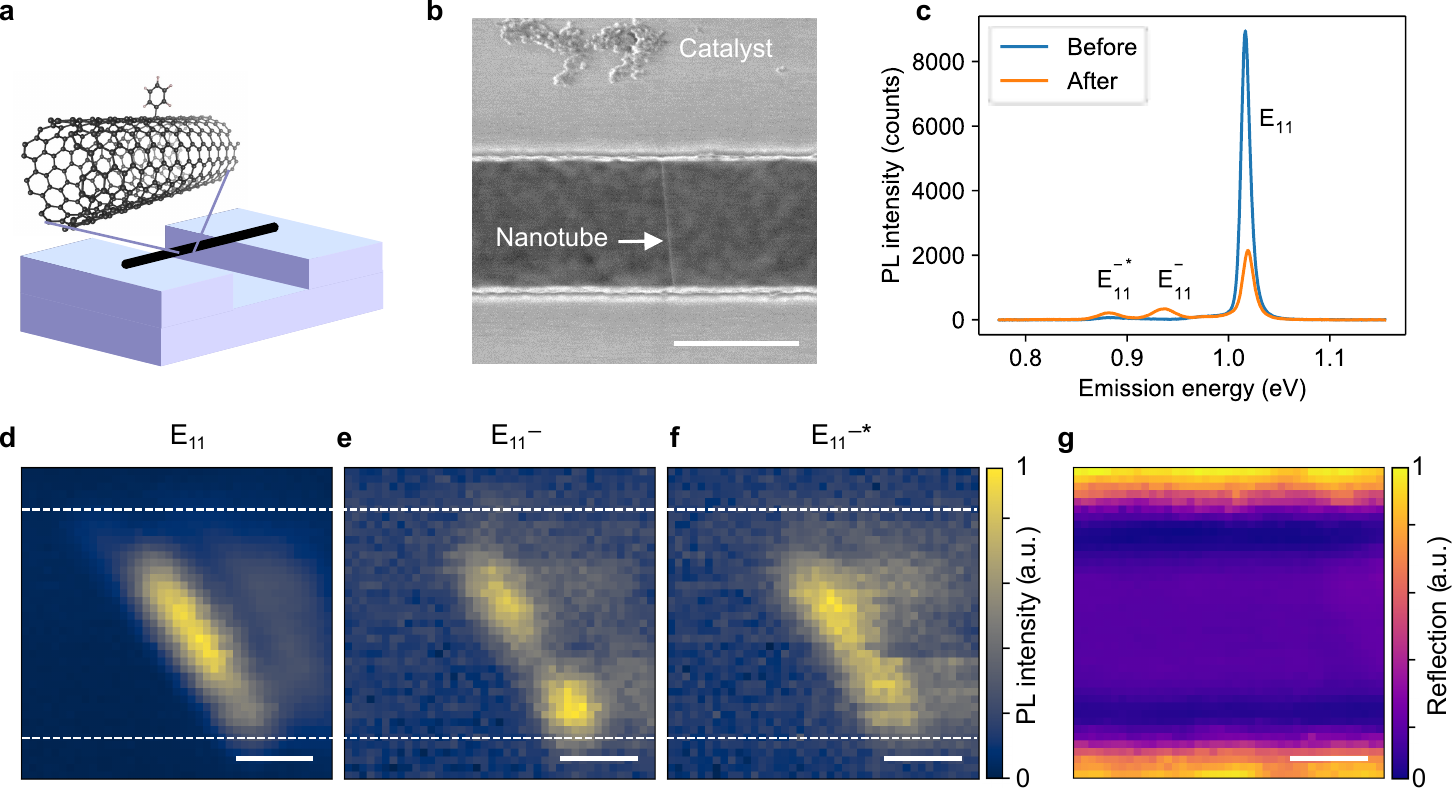}
	\caption{
		\label{fig1} Introducing organic color centers to air-suspended nanotubes using vapor-phase reaction. (a) A schematic of a functionalized SWCNT suspended across a trench on a Si substrate. (b) An scanning electron micrograph of a tube after the functionalization and the series of PL measurements. Particles on the top are patterned catalysts for growing SWCNTs and the nanotube is indicated by an arrow. (c) Representative PL spectra of an identical air-suspend (10,5) SWCNT before and after the functionalization taken with a laser power of 10~$\mu$W and an excitation energy of 1.59~eV. PL intensity maps of (d) E$_{11}$, (e) E$_{11}^-$, and (f) E$_{11}^{-*}$ emission from a (9,7) tube where the intensity is integrated within a window of 37.4, 32.5, and 28.5~meV centered at each emission peak, respectively. The white broken lines indicate the edges of the trench. (g) A reflection image in the same area, where  brighter and darker regions correspond to the surface of the substrate and the bottom of the trench, respectively. The scale bars in panels (b, d--g) are 1.0~$\mu$m. }
	\end{figure*}
	
	\section*{Results and Discussions}
	
	Air-suspended SWCNTs are grown across trenches on Si substrates by chemical vapor deposition \cite{Ishii:2019}, and vapor-phase reaction using iodobenzene is then conducted to create color centers. Figure \ref{fig1}a shows a schematic of a functionalized nanotube. Scanning electron microscopy confirms that the tubes stay suspended after the functionalization (Fig.~\ref{fig1}b). We emphasize that the vapor-phase reaction here differs from typical functionalization techniques established for dispersed SWCNTs in liquid \cite{Kim2018, Gifford2019a, Wu2018, Kwon2016a}. The solution process results in contaminating the tube surface and quenches PL due to interactions between SWCNTs and surrounding environment \cite{Lefebvre:2003}. It is noteworthy that directly immersing air-suspended tubes into water inevitably destroys the structures due to a high surface tension of the solvent (Fig.~S1). 
	
	We begin by examining PL spectra before and after the functionalization of a (10,5) SWCNT (Fig.~\ref{fig1}c) by using the coordinates of the tube on the chip \cite{Ishii:2019} to ensure that we are comparing the same individual tube. The pristine tube only shows exciton emission at a higher energy, whereas the functionalized tube shows two additional peaks at lower energies. We label the exciton emission at 1.02~eV as E$_{11}$ and the additional peaks at 0.94 and 0.88~eV as E$_{11}^-$ and E$_{11}^{-*}$, respectively. The lower energies of E$_{11}^-$ and E$_{11}^{-*}$ indicate that sp$^3$ defects of phenyl group are formed on SWCNTs which introduce dopant states \cite{Piao:2013}. No remarkable spectral shift of E$_{11}$ emission peak is detected, suggesting negligible changes in dielectric environment due to the vapor residue. The overall PL intensity reduces to less than a quarter which implies introduction of quenching sites in addition to color centers. Hereafter, we refer to defects that decrease the E$_{11}$ intensity as quenching sites and defects that give rise to the E$_{11}^-$ and the E$_{11}^{-*}$ peaks as color centers.
	
	Imaging measurements are performed to characterize the spatial distributions of E$_{11}$, E$_{11}^-$, and E$_{11}^{-*}$ emission. We scan over a (9,7) SWCNT to collect PL spectra and construct intensity maps for the three peaks by spectrally integrating intensities of each peak (Figs.~\ref{fig1}d--f). The edges of the trench can be identified using a reflection image in the same area (Fig.~\ref{fig1}g). Bright PL is emitted from the suspended region, as typically observed for air-suspended tubes \cite{Ishii:2019}. We find that luminescent profiles for E$_{11}^-$ and E$_{11}^{-*}$ are spatially overlapped with the profile for E$_{11}$, as expected for emission originating from color centers formed on the same tube. It is noted that the additional peaks E$_{11}^-$ and E$_{11}^{-*}$ show some spatial inhomogeneity in the intensity.

\begin{figure}
	\includegraphics{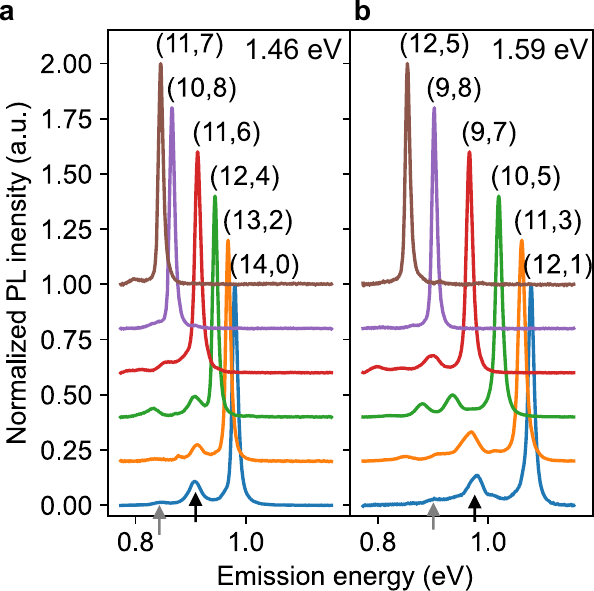}
	\caption{
		\label{fig2} Dopant state emission from various chiralities. PL spectra of functionalized SWCNTs collected with excitation laser energies of (a) 1.46 and (b) 1.59~eV and with an excitation power of 100~$\mu$W, where the spectra are displaced vertically for clarity. Black and gray arrows indicate E$_{11}^-$ and E$_{11}^{-*}$, respectively. Chirality $(n, m)$ is labeled next to the E$_{11}$ peaks. }
	\end{figure}

	E$_{11}^-$ and E$_{11}^{-*}$ emission from various chiralities are now studied by collecting PL spectra of more than 2000 individual SWCNTs. All PL data are obtained from a single substrate which assures that the reaction condition is the same, allowing for direct comparison among the chiralities. To acquire the data efficiently, we perform two sets of measurements with excitation energies of 1.46 and 1.59~eV which are near-resonant to many chiralities. Assuming that excitation is close to the E$_{22}$ energy, chiralities of SWCNTs are assigned based on the E$_{11}$ emission energy. We focus on 12 chiralities with sufficient numbers of tubes for statistical analysis, and typical spectra are shown in Figs.~\ref{fig2}a and \ref{fig2}b. Logscale PL spectra and PL excitation (PLE) maps for each chirality are presented in Figs.~S2 and S3, respectively. All chiralities exhibit E$_{11}^-$ and E$_{11}^{-*}$ emission except for (10,8), (11,7), and (12,5) SWCNTs whose E$_{11}^{-*}$ is beyond the low energy detection limit. We find that most E$_{11}^-$ peaks are taller than E$_{11}^{-*}$ peaks.

\begin{figure}
		\includegraphics{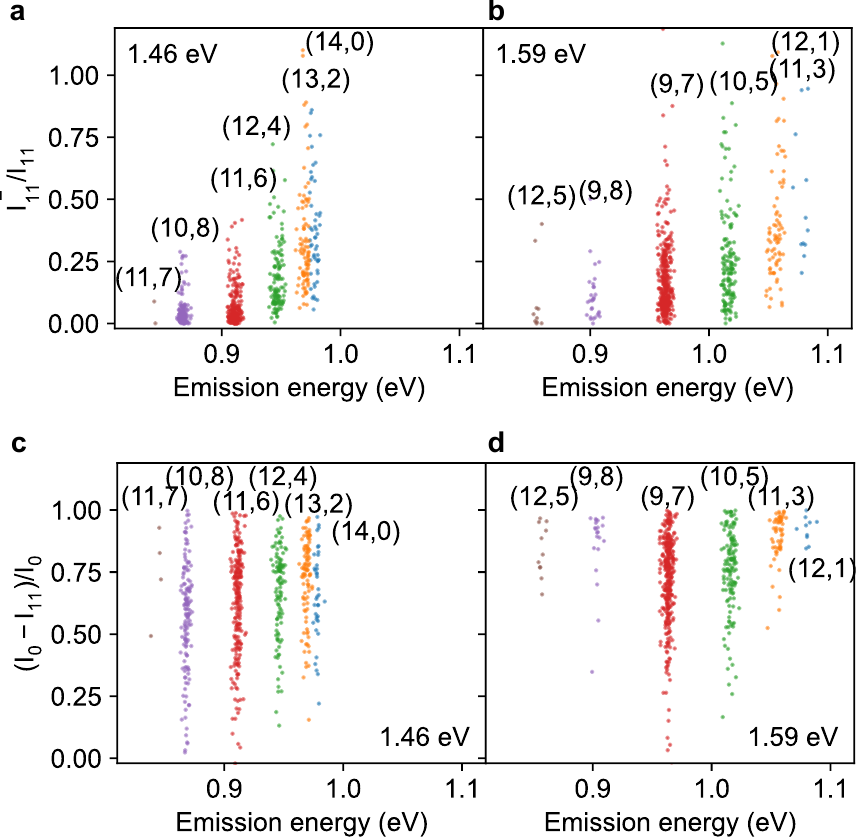}
		\caption{
			\label{fig3} Statistical analysis of PL intensities. Subpeak ratio as a function of emission energy for experiments conducted with excitation energies of (a) 1.46 and (b) 1.59~eV and an excitation power of 100~$\mu$W. Quenching degree measured with excitation energies of (c) 1.46 and (d) 1.59~eV and an excitation power of 10~$\mu$W.}
	\end{figure}

	Statistical analysis is performed by fitting a triple-Lorentzian function to PL spectra, and we first consider the intensity. In Figs.~\ref{fig3}a and \ref{fig3}b, subpeak ratio $I^{-}_{11}/I_{11}$ is plotted as a function of E$_{11}$ emission energy, where $I^{-}_{11}$ and $I_{11}$ are spectrally integrated intensities of E$_{11}^-$ and E$_{11}$ emission, respectively. The ratio can be regarded as a measure of the color center density, and we observe a monotonically increasing trend with emission energy. Although data dispersion is large, it indicates that smaller diameter tubes have more color centers.

	It is also possible to study the effects of the defects from the reduction of E$_{11}$ emission due to the functionalization. Figures \ref{fig3}c and \ref{fig3}d show quenching degree $(I_{0}-I_{11})/I_{0}$ where $I_0$ is spectrally integrated intensity of E$_{11}$ emission before the reaction. The quenching degree exhibits an increasing trend with emission energy as in the case of $I^{-}_{11}/I_{11}$.~The large variation of the ratio is likely caused by multiple factors including inhomogeneity among SWCNTs, temporal fluctuations in intensity (Fig.~S4), and positions of the defects \cite{Harrah:2011a}. We note that $(I_{0}-I_{11})/I_{0}$ reflects the effects of color centers in addition to quenching sites, as both results in reduced E$_{11}$ emission through trapping excitons.

	\begin{figure}
		\includegraphics{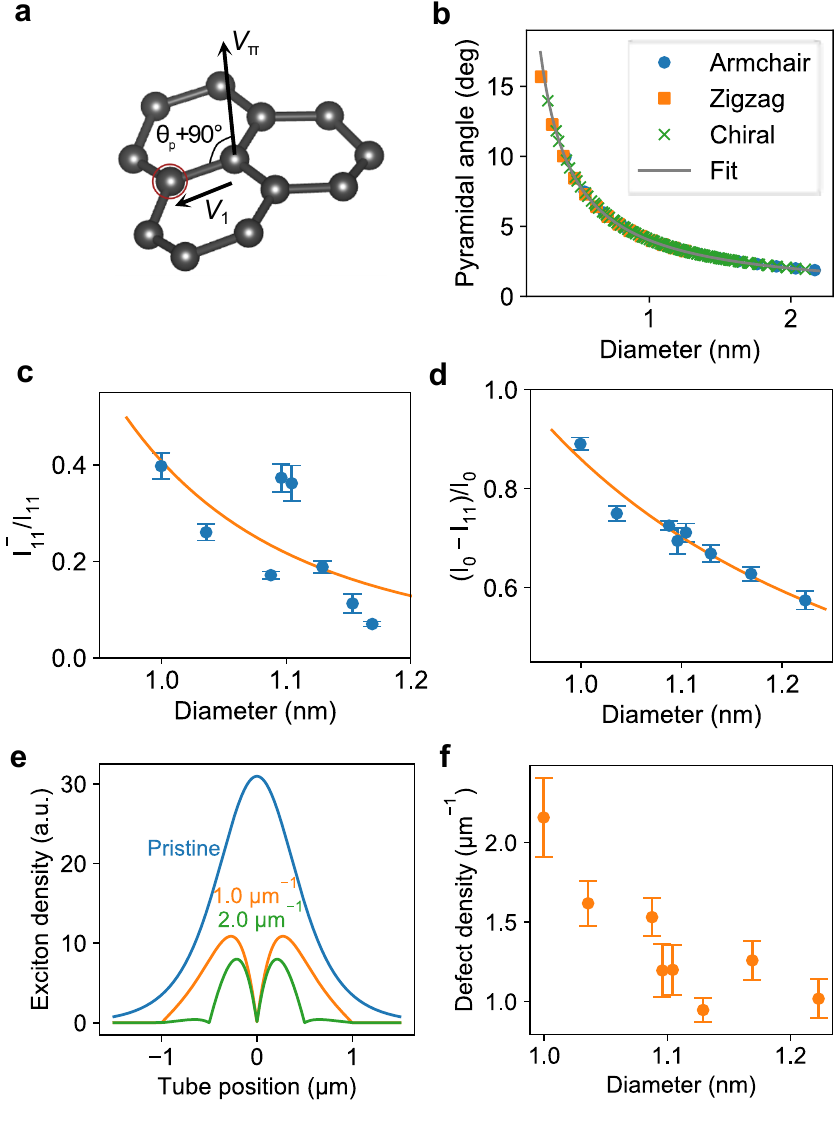}
		\caption{
			\label{fig4} Diameter-dependent reactivity. (a) A schematic defining the pyramidalization angle $\theta_p$ where $V_\pi$ is a $\pi$-orbital axis vector, $V_1$ is a unit vector pointing from the target atom to an adjacent atom \cite{Haddon2001}. The pyramidalization angle can be analytically estimated using a relationship $\cos (\theta _p+ \ang{90})=V_1 \cdot V_\pi$. (b) Diameter dependence of computed pyramidalization angle, where the angles of armchair, zigzag, and chiral tubes are separately plotted. Diameter dependence of (c) subpeak ratio and (d) quenching degree, where error bars are the standard error of the mean. The solid lines in the panels (b--d) are the fits. (e) Simulations of E$_{11}$ steady-state exciton density profile for no defects (blue), $\rho$ = 1.0 (orange), and $\rho$ = 2.0~$\mu$m$^{-1}$ (green). The origin of the coordinate system is taken to be the center of the tube. (f) Diameter dependence of estimated defect density, where error bars are the standard error of the mean. }
	\end{figure}	
		
	To quantitatively interpret the trend of the subpeak ratio and the quenching degree, a theoretical model is developed. Chemical reactivity of SWCNTs depends on both $\pi$-orbital pyramidalization angle \cite{Park2003} and $\pi$-orbital misalignment angle between adjacent pairs of conjugated C atoms \cite{Niyogi2002, Saha2018}. The former is subject to strain arising from the curvature of the tubular structure and is diameter dependent, whereas the latter originates from a bond angle with respect to the tube axis and is chiral angle dependent. As we observe a clear diameter dependence, our model considers the $\pi$-orbital pyramidalization angle $\theta_\mathrm{p}$ depicted in Fig.~\ref{fig4}a. The C-C bonds are more bent for larger $\theta_\mathrm{p}$, corresponding to larger strain. Figure \ref{fig4}b shows calculated pyramidalization angles as a function of the nanotube diameter $d$ along with a fit by a scaling law $\theta_\mathrm{p}=\delta/d$ where $\delta$ = 4.01$^\circ$~nm$^{-1}$ is the coefficient. Because the strain from the curvature increases the chemical reactivity \cite{Park2003}, we assume that the reduction in the activation energy is proportional to the pyramidalization angle. The activation energy of the reaction is then $E_\mathrm{a} (d)=E_\mathrm{a} (\infty)-\eta \theta_\mathrm{p} (d)$ where $E_\mathrm{a} (\infty)$ is the activation energy for graphene, and $\eta$ is the coefficient. According to the Arrhenius equation, it follows that the chemical reaction rates are proportional to
	\begin{equation}
		\exp \left( -\frac{E_\mathrm{a}(\infty)}{k_\mathrm{B}T}+\frac{1}{k_\mathrm{B}T} \cdot \frac{\eta \delta}{d} \right)
	\end{equation}
	 where $k_\mathrm{B}$ is the Boltzmann constant and $T$ = 298 K is the temperature. The ratios $I^{-}_{11}/I_{11}$ and $(I_{0}-I_{11})/I_{0}$ should therefore scale as $\exp \left(\frac{1}{k_\mathrm{B}T} \cdot \frac{\eta \delta}{d} \right)$.
	 
	 Since data dispersion is large, we use the average values of these ratios for each tube diameter (Figs.~\ref{fig4}c and \ref{fig4}d). It is confirmed that SWCNTs with smaller diameters show higher ratios, indicating that these tubes are more reactive despite the smaller surface areas. We fit the model to the experimental data, and both ratios show good agreement. The fit to $I^{-}_{11}/I_{11}$ yields $\eta = 44.3 \pm 10.0$~meV/deg, whereas the fit to $(I_{0}-I_{11})/I_{0}$ results in $\eta = 14.2 \pm 2.1$~meV/deg. The higher $\eta$ for $I^{-}_{11}/I_{11}$ by a factor of 3.12 indicates that the formation of color centers is more responsive to strain compared to quenching sites. We note that an opposite diameter dependence on chemical reactivity has been reported for a reaction with 4-hydroxybenzene diazonium \cite{Nair2007}, where electron transfer limits the reaction rate.

	 The ratio $(I_{0}-I_{11})/I_{0}$ also allows us to quantify the defect density. We start by modeling the exciton density profiles based on a steady-state one-dimensional diffusion equation 
	 \begin{equation}
	 	D \frac{d^2 n(z)}{dz^2} - \frac{n(z)}{\tau} + \frac{\Gamma _0}{\sqrt{2 \pi r^2}} \exp \left( -\frac{z^2}{2r^2} \right)=0
	 \end{equation}
	where $D$ is the diffusion coefficient, $n(z)$ is the E$_{11}$ exciton density, $z$ is the position on the tube, $\tau = 70$~ps is the intrinsic lifetime of excitons \cite{Ishii:2019}, $\Gamma_0$ is the exciton generation rate, and $r = 530$~nm is the $1/e^2$ radius of the laser spot. The first term accounts for the exciton diffusion, the second term represents the intrinsic recombination, and the third term describes exciton generation which is proportional to the Gaussian laser profile. We consider SWCNTs with infinite length and set the boundary conditions to be $n(\pm \infty)=0$. Additional boundary conditions $n(z_\mathrm{d})=0$ are imposed for the functionalized tube where $z_\mathrm{d}$ is the position of defects, assuming that the sites are uniformly distributed with density $\rho$. For the diffusion coefficient, we use the expression $D=D_0 (d/d_0)^\alpha$ where $D_0 = 15.36~\mathrm{cm}^2/$s is the diffusion coefficient at diameter $d_0 = 1.00$~nm and $\alpha = 2.56$ is the exponent \cite{Ishii:2019}. The diffusion equation is numerically solved to obtain $n(z)$ and the results are plotted for various site densities in Figs.~\ref{fig4}e and S5a. When the defect separation is much shorter than the laser spot diameter, the quenching process becomes dominant over the intrinsic decay and results in a significant decrease in the exciton density.
	
	The defect density is estimated by comparing $(I_{0}-I_{11})/I_{0}$ obtained from the experiments with the simulations. We use the experimental data of tubes on the widest trenches with 3.0~$\mu$m widths, and the PL intensity in the simulation is computed by integrating $n(z)$ (Fig.~S5b). The only unknown parameter $\rho$ is extracted by matching the simulated quenching degree with the experimental values. The results are plotted as a function of the diameter in Fig.~\ref{fig4}f. The defect density ranges from 0.95 to 2.2~$\mu$m$^{-1}$ and shows a diameter dependence which is consistent with the trend of the reactivity. It is worth mentioning that the estimated defect density includes contributions from color centers and quenching sites as they both reduce the number of E$_{11}$ excitons.
		
	\begin{figure}
		\includegraphics{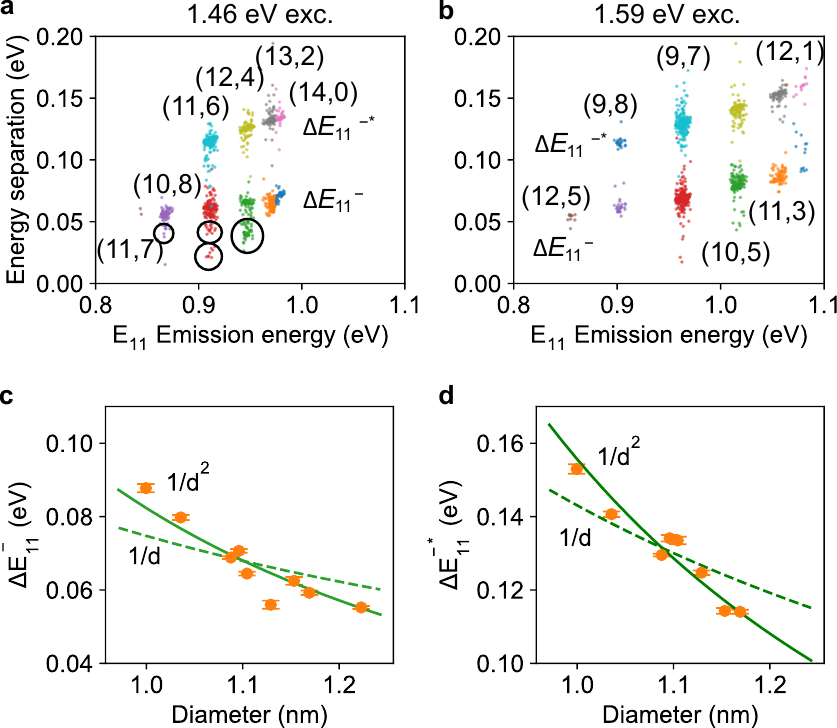}
		\caption{
			\label{fig5}  Statistical analysis of emission energies. Energy separation $\Delta E_{11}^{-*}$ and $\Delta E_{11}^-$ of functionalized SWCNTs as a function of emission energy, where the data are collected with excitation energies of (a) 1.46 and (b) 1.59~eV and a power of 100~$\mu$W. The color of the dots represents the nanotube chirality and isolated clusters are indicated by circles. Diameter dependence of average (c) $\Delta E_{11}^-$ and (d) $\Delta E_{11}^{-*}$, where error bars are the standard error of the mean. The solid and broken lines are the fits by the power laws $1/d^2$ and $1/d$, respectively. The data are better described by ${\Delta E_{11}^-=A}^-/d^2$ and ${\Delta E_{11}^{-*}=A}^{-*}/d^2$, where $A^- = 84.0 \pm 2.64$~meV$\cdot$nm$^2$ and $A^{-*} = 154 \pm 1.43$~meV$\cdot$nm$^2$ are the coefficients for $\Delta E_{11}^-$ and $\Delta E_{11}^{-*}$, respectively.}
	\end{figure}	

	We now proceed to analyze E$_{11}^-$ and E$_{11}^{-*}$ peak positions. Energy separations ${\Delta E_{11}^-=E}_{11}-E_{11}^-$ and $\Delta E_{11}^{-*}=E_{11}-E_{11}^{-*}$ can be interpreted as trapping potential depths for E$_{11}^-$ and E$_{11}^{-*}$ excitons, respectively, and we plot $\Delta E_{11}^-$ and $\Delta E_{11}^{-*}$ as a function of $E_{11}$ in Figs.~\ref{fig5}a and \ref{fig5}b. The energy separations show correlated increase with E$_{11}$, confirming that E$_{11}^-$ and E$_{11}^{-*}$ originate from dopant states of E$_{11}$ exciton, and not of other states such as E$_{22}$ and E$_{33}$ excitons. It is noted that we observe smaller clusters for $\Delta E_{11}^-$ as marked by circles in Fig.~\ref{fig5}a, whose trapping potentials are smaller than the main clusters. The differences in E$_{11}^-$ could be assigned to different binding configurations of the phenyl functional group, where ortho- and para-configurations exhibit different emission energies \cite{Saha2018,Gifford2019a,He2017}. We similarly interpret the E$_{11}^{-*}$ emission to be arising from other binding configurations. 

	The diameter dependence of the trapping potentials provides additional insight to the nature of the dopant states. The average values of $\Delta E_{11}^-$ and $\Delta E_{11}^{-*}$ for each chirality are plotted as a function of the diameter in Fig \ref{fig5}c and 5d. The dependence shows a monotonic decrease and differs from the constant energy separation of 130~meV  for the $K$-momentum excitons, and we thus exclude them from the origin of E$_{11}^-$ and E$_{11}^{-*}$. To describe the dependence, we consider the $1/d$ scaling observed for the exciton binding energies and the $1/d^2$ scaling for singlet-triplet splitting \cite{Capaz:2006,Santos:2011,Nagatsu:2010, Matsunaga:2010}. The $1/d^2$ scaling yields better fits to both $\Delta E_{11}^-$ and $\Delta E_{11}^{-*}$ than the $1/d$ scaling.

	The values of the energy separations $\Delta E_{11}^-$ and $\Delta E_{11}^{-*}$ observed in this work show more similarity to triplet and trion states. Examining the higher energy peak E$_{11}^-$, $\Delta E_{11}^-$ = 84.0~meV at $d$ = 1~nm is as high as the energy splitting reported for triplet excitons; Matsunaga \textit{et al.} found $1/d^2$ dependence of energy separation for laser-induced defects in air-suspended SWCNTs \cite{Matsunaga:2010}, with a value of 70~meV at $d$ = 1~nm.  Nagatsu \textit{et al.} showed the $1/d^2$-dependent energy separation in air-suspended SWCNTs for H$_2$-adsorption-induced peaks where $\Delta E_{11}^-$ = 68~meV at $d$ = 1~nm is attributed to triplet excitons \cite{Nagatsu:2010}.  The comparable values of the energies and the $1/d^2$ scaling for E$_{11}^-$ suggest that the triplet exciton state is brightened at the color centers. Further study with optically detected magnetic resonance \cite{Stich:2014} and magneto-PL spectroscopy \cite{Kwon2019} would be required to clarify the triplet origin of E$_{11}^-$ excitons. Considering the lower energy peak E$_{11}^{-*}$, $\Delta E_{11}^{-*}$ = 154~meV at $d$ = 1~nm is close to 175~meV-separation between exciton and trion energies for air-suspended SWCNTs \cite{Yoshida:2016} with a diameter of 1~nm. Using chemical \cite{Kwon2019} or electric-field \cite{Gluckert2018} doping to investigate trions trapped at color centers in air-suspended tubes may help elucidate the origin of the state. We note that the tubes studied here have larger diameters than typical SWCNTs dispersed in liquid \cite{Gifford2019a,Saha2018}, but the observed $\Delta E_{11}^-$ and $\Delta E_{11}^{-*}$ are consistent with the extrapolation of the $d$ dependence for smaller diameters.		

	\begin{figure}
		\includegraphics[scale=1.05]{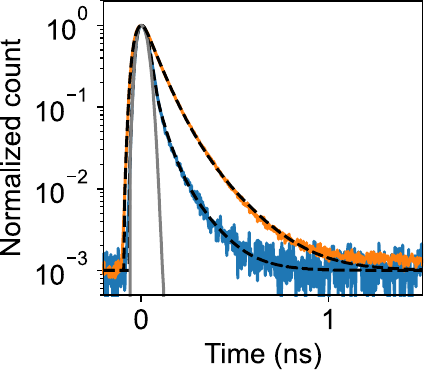}
		\caption{
			\label{fig6} Time-resolved PL properties. PL decay curves of E$_{11}$ emission (blue) and E$_{11}^-$ and E$_{11}^{-*}$ emission (orange) from a functionalized (11,3) SWCNT suspended across a trench with a width of 1.0~$\mu$m measured by an excitation energy of 1.59~eV and a power of 10 nW using the pulsed laser. The broken lines are the fits by a biexponential decay function convoluted with the instrument response function (gray), showing $t_1$ = 42.2 and $t_2$ = 122.0~ps for E$_{11}$ and $t_1$ = 69.1 and $t_2$ = 172.0~ps for E$_{11}^-$ and E$_{11}^{-*}$.}
	\end{figure}

	The dynamics of the color center emission is also investigated by time-resolved PL measurements to compare to the decay lifetimes of E$_{11}$. We use a pulsed laser for excitation and the E$_{11}$ emission is spectrally differentiated from the E$_{11}^-$ and E$_{11}^{-*}$ emission with a band-pass filter and a long-pass filter (Fig.~S6). Figure \ref{fig6} shows PL decay curves for these emission peaks from an (11,3) functionalized tube. Decay lifetimes are extracted by fitting a biexponential function $a_1\exp{\left(-t/\tau_1\right)+a_2\exp{\left(-t/\tau_2\right)}}$ convoluted with the instrument response function, where $\tau$ is the decay lifetime and $a$ is the amplitude with the subscripts 1 and 2 denoting fast and slow components, respectively. The fast component is assigned to the decay of bright excitons whereas the slow component reflects the dynamics of dark excitons \cite{Ishii:2019}. The fast decay for E$_{11}^-$ and E$_{11}^{-*}$ emission exhibits lifetime longer by 1.64 times compared to E$_{11}$, as observed for dopant states in solution-processed tubes \cite{Miyauchi:2013,Hartmann:2016}.

	In summary, we have demonstrated functionalization of air-suspended SWCNTs using iodobenzene as a precursor. The comparison of PL spectra before and after the vapor-phase reaction shows additional peaks E$_{11}^-$ and E$_{11}^{-*}$ from color centers and PL intensity reduction of the E$_{11}$ peaks. Twelve representative chiralities are characterized using spectra from more than 2000 individual tubes, where the diameter dependent subpeak ratio and quenching degree are observed. We have modeled the diameter dependent reactivity which is found to be proportional to $\exp(1/d)$, explaining the experimental results. By further performing the exciton diffusion simulations, we have estimated the defect density and found that these values are also diameter dependent. The analysis of peak energies reveals that both E$_{11}^-$ and E$_{11}^{-*}$ states originate from dopant states of E$_{11}$ excitons and have trapping potentials scaling as $1/d^2$. We observe a longer PL lifetime for dopant states, similar to the reports on solution-processed tubes. By elucidating the exciton physics as well as functionalization chemistry, color centers in air-suspended SWCNTs should provide new opportunities in photonics and optoelectronics for quantum technologies.
	
	\section*{Materials and Methods}
	Electron-beam lithography and dry etching are used to fabricate trenches on Si substrates \cite{Ishii:2017} with a depth of $\sim$1~$\mu$m and a width of up to 3.0~$\mu$m. Another electron-beam lithography is conducted to define catalyst areas near trenches, and Fe-silica catalyst dispersed in ethanol are spin-coated and lifted off. SWCNTs are synthesized over trenches using alcohol chemical vapor deposition \cite{Ishii:2017, Ishii:2019} under a flow of ethanol with a carrier gas of Ar/H$_2$ at 800\si{\celsius} for 1~min.
	
	Vapor-phase reaction is used to functionalize air-suspended nanotubes with iodobenzene as a precursor. As-grown SWCNTs on the Si substrates are placed facing up inside a glass chamber having a diameter of 15~mm and a height of 5~mm. Iodobenzene (5 $\mu$L) is introduced to the bottom of the chamber by a micropipette. The chamber is then covered with a quartz slide and sealed using high vacuum grease. We leave the chamber for 10~min to fill it with iodobenzene vapor, after which the reaction is triggered by irradiating the sample with 254-nm UV light through the quartz slide. The samples are collected from the chamber and stored in dark for characterization by subsequent spectroscopy. 
	
	PL spectra are obtained with a home-built scanning confocal microscope \cite{Ishii:2019}, where we use a continuous-wave Ti:sapphire laser for excitation and a liquid-N$_2$-cooled InGaAs photodiode array attached to a 30-cm spectrometer for detection. Laser polarization is kept perpendicular to trenches, and the beam is focused using an objective lens with a numerical aperture of 0.85 and a focal length of 1.8~mm. The $1/e^2$ diameters of the focused beams are 1.31 and 1.06~$\mu$m for excitation energies of 1.46 and 1.59~eV, respectively, where the diameters are characterized by performing PL line scans perpendicular to a suspended tube. PLE spectroscopy is conducted by scanning excitation wavelength at a constant power \cite{Ishii:2015}. The reflected laser light is collected with a biased Si photodiode for reflection images. 
	
	For time-resolved PL measurements, the Ti:sapphire laser is switched from continuous wave to $\sim$100-fs pulses with a repetition rate of 76~MHz. A fiber-coupled superconducting single-photon detector is used to measure PL decay. Emission from E$_{11}$ excitons and dopant states are separately obtained with a band pass filter and a long pass filter.
	
	\section*{Acknowledgments}
	This work is supported in part by MIC (SCOPE 191503001), JSPS (JP18H05329, JP20H00220, JP20H02558, JP20K15112, JP20K15137), MEXT (Nanotechnology Platform JPMXP09F19UT0077), NSF (RAISE-TAQS PHY-1839165), and JST (CREST JPMJCR20B5). D.K. acknowledges support from RIKEN Special Postdoctoral Researcher Program. K.O. is supported by JSPS Research Fellowship. We thank the Advanced Manufacturing Support Team at RIKEN for technical assistance. 

	\section*{Author contributions}
	Y.K.K., Y.H.W., and S.M. conceptualized the idea, and formulated the overarching research goals. A.I., T.I., and R.X. synthesized SWCNTs. X.W. and J.F. conducted the functionalization. D.K., A.I., and K.O. performed the optical measurements. D.K., A.I., and Y.K.K. interpreted the results. D.K. and Y.K.K. wrote the original draft with input from all the authors.

	\end{document}